\documentclass[10pt,aps,prb,twocolumn,superscriptaddress,floatfix,showpacs,longbibliography]{revtex4-1}
\usepackage[utf8]{inputenc}
\usepackage[T1]{fontenc}
\usepackage{graphicx}
\usepackage{tabularx}
\usepackage[usenames,dvipsnames,table]{xcolor}
\usepackage[version=3]{mhchem}
\usepackage{braket}
\usepackage{upgreek}
\usepackage{hyperref}
\usepackage{amsmath, amssymb}
\usepackage[normalem]{ulem}
\usepackage{soul}
\usepackage{bbold}

\definecolor{mark}{rgb}{0.85, 0.9, 1}

\hypersetup{hidelinks}
\sethlcolor{mark}

\begin{document}

\title{Brute-force positivization of $J_1 - J_{2}$ model ground states}

\author{P. A. Bannykh}
\affiliation{Theoretical Physics and Applied Mathematics Department, Ural Federal University, Ekaterinburg 620002, Russia}
\author{O. M. Sotnikov}
\affiliation{Theoretical Physics and Applied Mathematics Department, Ural Federal University, Ekaterinburg 620002, Russia}
\affiliation{Russian Quantum Center, Skolkovo, Moscow 121205, Russia}
\author{V. V. Mazurenko}
\affiliation{Theoretical Physics and Applied Mathematics Department, Ural Federal University, Ekaterinburg 620002, Russia}
\affiliation{Russian Quantum Center, Skolkovo, Moscow 121205, Russia}

\begin{abstract}
Exploring sign structures of quantum wave functions attracts considerable attention due to the potential for advances in modeling complex phases of matter. This stimulates developing different optimization procedures for imitating and manipulating sign structures of quantum states. In this work, utilizing a brute force approach based on a set of single-qubit transformations we evaluate protocols enabling positivization of the one-dimensional $J_1 -J_2$ model ground states in the regime of strong frustration. Based on the obtained positivization results, we show the difference between the cases of periodic and open boundary conditions, and also establish the dependence of the sign structure on parity of the simulated spin chains.
\end{abstract}

\maketitle

\section{Introduction}

An $N$-qubit wave function with real-valued amplitudes in a given computational basis can be characterized by one of $2^{2^N}$ possible sign structures. Being intimately related to quantum correlations \cite{Fisher}, sign structure tuning can be used for manipulation of wave function entanglement \cite{Dicke_sign}, which is in demand in quantum computing. Switching between different sign configurations can be realized with combinations of one-qubit Pauli-Z, two-qubit controlled-Z (CZ) and multi-qubit controlled-$Z$ (CC$\ldots$CZ) gates that change the sign for particular basis states or its groups \cite{Nielsen}. For instance, the Pauli-Z gate acting on a particular qubit alters the sign of all basis states for which the corresponding qubit has the $\ket{1}$ state. In turn, CZ gate entangling a pair of qubits affects all the basis states having $\ket{1}$ for both qubits. However, due to an astronomically large number of the sign structures, search for a particular gate combination transforming a given sign configuration to a target one is an open problem whose solution is known only in a limited number of cases. 

In condensed matter physics, a related task is known as positivization \cite{Torlai_positivization}. Here one attempts to find unitary transformations that reduce the number of basis states with negative amplitudes while preserving their absolute values. Put simply, the closer a wave function is to the positive-amplitude form, the easier it is to model such a quantum state by using classical resources. As a supporting example one can consider constructing classical representations of the quantum states with classical neural networks \cite{Carleo_Troyer, Clark, Dicke_RBM} for exploring properties of quantum many-body systems. In the case of highly frustrated magnetic systems with competing interactions the authors of Ref.\onlinecite{Bagrov1} have shown that learning sign structure of a quantum state  with a neural network is generally much more difficult than learning amplitudes. As pointed out in Ref.\onlinecite{Choo}, optimizing a neural network toward the ground state is extremely challenging if an appropriate sign structure is not imposed. To facilitate learning of sign structure of wave functions different strategies \cite{Neupert, Ou, Szabo, Bagrov2} have been developed. 

Another way to ease finding ground state of non-stoquastic systems whose ground states are characterized by non-trivial sign configurations is to transform the target Hamiltonian by using insider information about properties of underlying quantum state. Historically, a unitary transformation derived by Marshall \cite{Marshall} with assistance of Peierls for a two-sublattice antiferromagnet is the first documented example of performing positivization procedure. Being exact in the limit of the antiferromegnetic Heisenberg model with only nearest-neighbour interactions defined on one-dimensional or two-dimensional square lattice, this approach is likewise used in simulations of more complicated spin systems that are characterized by couplings between more distant spins \cite{Choo}. As a prominent example, we consider the one-dimensional $J_1 - J_2$ model with antiferromagnetic interactions between nearest and next-nearest neighbours for which exact diagonalization calculations reported in Refs.\onlinecite{Torlai2018, Becca} revealed a validity of Marshall-Peierls rule in the range $J_2 \in [0, 0.5]$. This considerably improves the performance of neural network approaches \cite{Shamim, Becca} used for imitating the ground state of such a model and stimulates search for other unitary transformations pozitivizing ground state solutions in a wider parameter range, $J_2 >0.5$. 

In our paper we report on a systematic study of the sign structure of the one-dimensional $J_1 - J_2$ model ground states. The model we consider is defined on finite spin chains with open and periodic boundary conditions, since results for both cases can be found in the literature. Our focus is on the positivization of the ground state solutions with single-qubit rotational $R^z (\theta)$ gates. For that we use brute force approach by quantifying sign structures of all possible combinations of these gates with $\theta = \pi$ and $\pi/2$. Despite of the fact that all the spin chains we consider are characterized by even number of spins ($N$ = 6, 8, 10, .. 30), we have found that the details of the positivization schemes depend on the parity of $N_{\frac{1}{2}} = N/2$ value. We also examine combinations of one- and two-qubit gates and show that they allow to achieve better positivization results in the regime of strong frustration. Additionally for the developed positivization protocols we derive the parent Hamiltonians. 

\section{Spin model and methods}
\begin{figure}[t]
  \centering
  \includegraphics[width=0.93\linewidth]{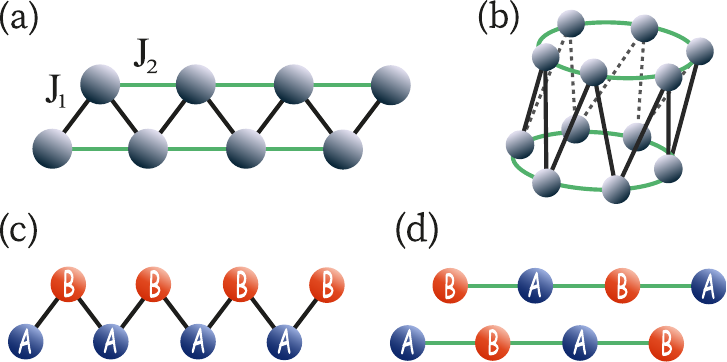}
  \caption{Schematic of the one-dimensional $J_1-J_2$ model with open (a) and periodic (b) boundary conditions. Two variants of subsystem partition for the limiting cases of $J_2=0$ (c) and $J_1=0$ (d). The black and green lines denote the magnetic interactions between nearest ($J_1$) and next-nearest ($J_2$) neighbours, respectively.}
  \label{fig1}
\end{figure}

In this paper we consider one-dimensional $J_1 - J_2$ Heisenberg model with antiferromagnetic interactions which is defined with the following Hamiltonian  
\begin{equation}\label{Hamiltonian}
\hat{H} = J_1 \sum\limits_{\braket{i,j}} \hat{\mathbf{S}}_i \cdot \hat{\mathbf{S}}_j + J_2 \sum\limits_{\braket{\braket{i,j}}} \hat{\mathbf{S}}_i \cdot \hat{\mathbf{S}}_j.
\end{equation}
Here $J_1$ ($J_2$) is the exchange interaction between nearest (next-nearest) spins in the chain (Fig.\ref{fig1}) and $\hat{\mathbf{S}}_i = (\hat{S}^x_i, \hat{S}^y_i, \hat{S}^z_i)$ is the spin-1/2 operator acting on the $i$-th spin, where $\hat{S}^x_i, \hat{S}^y_i, \hat{S}^z_i$ represent the corresponding projections. 

Previous numerical studies have clarified basic properties of the one-dimensional $J_1 - J_2$ model with antiferromagnetic interactions in a wide range of parameters. For instance, the authors of Refs.\onlinecite{Majumdar_I, Majumdar_II} have explored the model on short chains from 3 to 10 spins with periodic boundary conditions and revealed that at $J_{2} / J_{1} = 0.5$ (the Majumdar-Ghosh point) the ground state is doubly degenerate and can be represented as the tensor product of the singlet states. 
In turn, exact diagonalization of the model with up to 20 spins \cite{tonegawa} has demonstrated that there is a transition between finite singlet-triplet gap and gapless phases. Ref.\onlinecite{Shu} reports on reconstructing quantum phase diagram of the $J_1 -J_2$ model with calculating fidelities for the excited states on the basis of small-size chain results obtained by using exact diagonalization. The description of the spin gap dependence on $J_2/J_1$ ratio was one of the main objectives in the study \cite{White} based on the density matrix renormalization group (DMRG) approach. Besides the authors of this work have explored spin-spin correlations and revealed the dimerization in a certain range of next-nearest neighbour coupling. Since in one-dimensional case DMRG enables simulating very long systems (with up to 400 spins as reported in Ref.\onlinecite{White}), these results are traditionally employed when benchmarking numerical schemes developed for solving many-body problems, for instance, those based on neural networks \cite{transformer}.  

In our study we explore chains of size ranging from 6 to 30 spins with open and periodic boundary conditions schematically visualized in Fig.\ref{fig1} (a) and (b). We set $J_1 = 1$ and $J_2$ is varied in the range $[0, 2]$. To find the ground states of the considered $J_1-J_2$ model we use the exact diagonalization approach realized in the SpinED package \cite{SpinED}. 

\begin{figure*}[ht]
  \centering
  \includegraphics[width=0.85\linewidth]{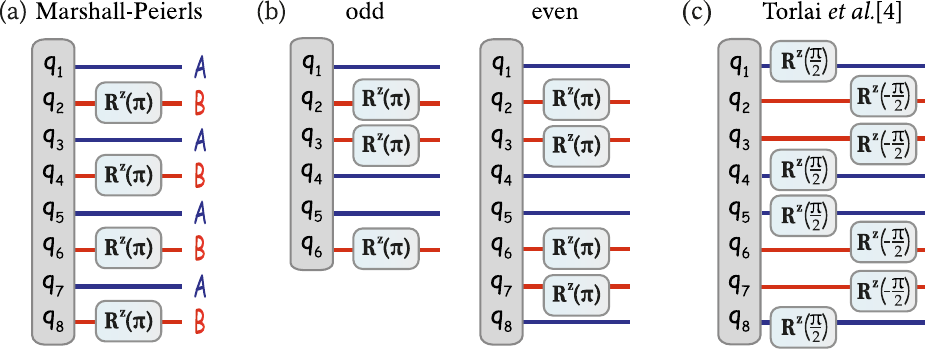}
  \caption{Schematic of one-qubit gates protocols used in this work for positivization of the ground states of the one-dimensional $J_1-J_2$ model. ${\rm R}^{z} (\theta)$ denotes the rotation operator about the z axis by $\theta$. (a) Marshall-Peierls scheme derived in Ref.\onlinecite{Marshall} for the case of $J_{2} = 0$. (b) Odd and even protocols obtained in this work with brute-force positivization procedure for quantum systems characterized by different parity of $N_{\frac{1}{2}}$. (c) The protocol introduced in Ref.\onlinecite{Torlai_positivization} by using an optimization procedure in the case of the one-dimensional $J_1 - J_2$ model with periodic boundary conditions at $J_2 / J_1 = 2$ for even $N_{\frac{1}{2}}$. The blue line marks the spins belonging to the sublattice A, the red line \--- B.}
  \label{fig2}
\end{figure*}

To characterize the sign structures of the calculated ground states $\ket{\Psi} = \sum_{\boldsymbol \sigma} \Psi (\boldsymbol \sigma) \ket{\boldsymbol \sigma}$ with real-valued amplitudes we use the following measure introduced in Ref.\onlinecite{Torlai2018}:
\begin{equation}
\braket{\rm Sign}_{\psi} = \sum_{\boldsymbol \sigma} {\rm sign}(\Psi(\boldsymbol \sigma))\cdot|\Psi (\boldsymbol \sigma)|^2,
    \label{metric}
\end{equation}
where $\Psi(\boldsymbol \sigma)$ represents the probability amplitudes of the $\ket{\boldsymbol \sigma}$ basis states. The values of $\braket{\rm Sign}_{\psi}$ for raw eigenstates of the $J_1 -J_2$ model obtained from exact diagonalization are close to or exactly zero, which evidences on a balance between basis states with positive and negative amplitudes. 

The manual positivization of the calculated ground states for small-size systems from 6 to 18 spins was performed by using a unitary transformation $\hat U = \otimes_{i=1}^{N} \hat U_{i}$, for which the choice of the single-qubit operators is based on the results obtained for one-dimensional $J_1 - J_2$ model in the previous works. More specifically, in the limit $J_2 = 0$ the ground state of Eq.\ref{Hamiltonian} can be represented as \cite{Marshall} 
\begin{eqnarray}
\ket{\Psi} = \sum_{\boldsymbol \sigma} (-1)^{N_B (\boldsymbol \sigma)} \Psi_p (\boldsymbol \sigma) \ket{\boldsymbol \sigma},
\end{eqnarray}
where $\Psi_{p} (\boldsymbol \sigma)$ is the positive amplitude, $N_{B} (\boldsymbol \sigma)$ is the number of spins that belong to the B sublattice (Fig.\ref{fig1} c) and have the $\ket{\uparrow}$ state.
This is nothing, but the famous Marshall-Peierls rule\cite{Marshall} (MPR) for antiferromagnets on bipartite lattices with nearest-neighbor interactions. On the other hand, the connection between the initial wave function $\ket{\Psi}$ and its positivized analog $\ket{\Psi_{p}}$ is defined through $\hat U_{\rm MPR} = \otimes_{j \in B} e^{i \pi \hat \sigma^z_{j} /2}$ that is the tensor product of rotation operators, $R^{z}_{j} (\pi)$ acting on the spins from the B sublattice about z axis, $\ket{\Psi_p} = \hat U_{\rm MPR} \ket{\Psi} $. In the opposite limit $J_1 = 0$ for a system with open and periodic boundary conditions, when N/2 is even, the exact positivization can be done. In this case one performs the MPR transformation after sorting the spins in each chain with respect to the sublattices A and B as visualized in Fig.\ref{fig1} d. In agreement with these results, the authors of Ref.\onlinecite{Torlai_positivization} have explored the ground state at $J_2 = 2$ and revealed a similar positivization protocol with $R^z (\frac{\pi}{2})$  and $R^z (-\frac{\pi}{2})$ for spins from the A and B sublattices, respectively.

Taking into account these results of the previous works it is convenient to start with thus found transformations for the limiting cases and use them to explore the system at intermediate values of $J_2$ in the range from 0 to 2. Thus, the set of the single-qubit unitary transformations we use for brute-force positivization includes identity, $I$ and rotational operators, $R^z(\pm\pi)$ and $R^z (\pm\frac{\pi}{2})$. According to the employed procedure for a $N$-qubit system one is to consider $5^N$ combinations (circuits) of one-qubit gates, prepare the corresponding wave functions and estimate the value of the sign average, Eq.\ref{metric}. For each $J_2$ we are interested in finding a combination of gates which produces the state with the largest absolute value of $\braket{\rm Sign}_{\psi}$.     

\section{Results}
\subsection{Open boundary conditions}
\begin{figure}[!b]
  \centering
  \includegraphics[width=0.93\linewidth]{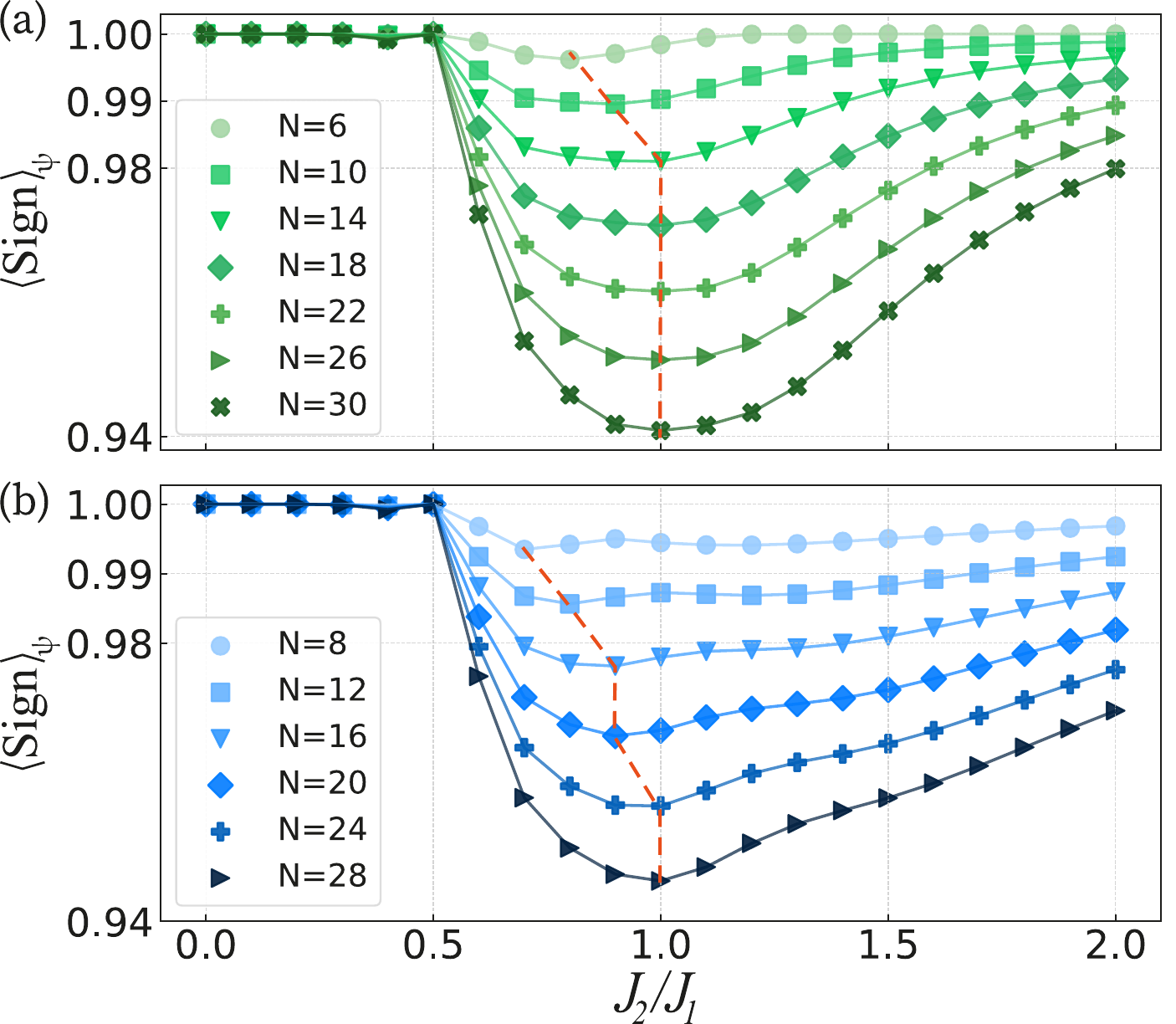}
  \caption{Values of sign function, Eq.\ref{metric} calculated for the pozitivized ground states of the $J_1-J_2$ model with odd (a) and even (b) $N_{\frac{1}{2}}$.  The red line indicates the shift of the minimum value of the sign function with the increase of the system size.}
  \label{fig3}
\end{figure}

We start with discussing the results obtained for the systems with open boundary conditions (OBC). Figures \ref{fig2} (a) and (b) demonstrate the positivization protocols found with manual procedure described in the previous section. The corresponding values of the sign function, Eq.\ref{metric} are presented in Fig.\ref{fig3}. We have found that the protocol presented in Fig.\ref{fig2} a, that corresponds to the Marshall-Peierls rule, provides maximal values of $\braket{\rm Sign}_{\psi}$ in the range $J_2 \in [0,0.5]$. In turn, in the range $J_2 \in [0.5, 2]$ the best positivization is achieved with either odd or even  schemes depending on parity of $N_{\frac{1}{2}} = N/2$ depicted in Fig.\ref{fig2} (b). The basic difference between Marshall-Peierls and odd (even) protocols is in the way of partitioning the whole system into two sublattices A and B, and applying rotational gates. While the former scheme corresponds to the ABABAB$\ldots$AB bipartition (Fig.\ref{fig1} (c)), the latter  reproduces the ABBAABBA decomposition that is visualized in Fig.\ref{fig1} (d) and takes place in the limit $J_2 \rightarrow \infty$ as discussed in the previous section. 

At $J_2/J_1 = 0.5$ which is known as the Majumdar-Ghosh point, both Marshall-Peierls and odd (even) protocols provide exact positivization of the ground state. At this point, the ground wave function of the $N$-spin system is nothing but the tensor product of the singlet states \cite{Majumdar_I, Majumdar_II}
\begin{eqnarray}
\ket{\Psi_{\rm MG}} = \ket{\Psi_{\rm S}}_{1,2}  \otimes \ket{\Psi_{\rm S}}_{3,4} \otimes ...\otimes \ket{\Psi_{\rm S}}_{N-1, N},
\end{eqnarray}
where $\ket{\Psi_{\rm S}}_{i,j} = \frac{1}{\sqrt{2}} (\ket{\uparrow \downarrow}_{i,j} - \ket{\downarrow \uparrow}_{i,j})$. Such a state can be fully positivized with the Marshall-Peierls protocol (Fig.\ref{fig2}), which provides the following transformation \cite{Torlai2018} $\ket{\Psi_{\rm S}}_{i,j} \rightarrow \ket{\Psi_{\rm T}}_{i,j} = \frac{1}{\sqrt{2}} (\ket{\uparrow \downarrow}_{i,j} + \ket{\downarrow \uparrow}_{i,j})$. The same positivization result can be reproduced with the odd (even) protocols.  

\begin{figure}[!t]
  \centering
  \includegraphics[width=0.93\linewidth]{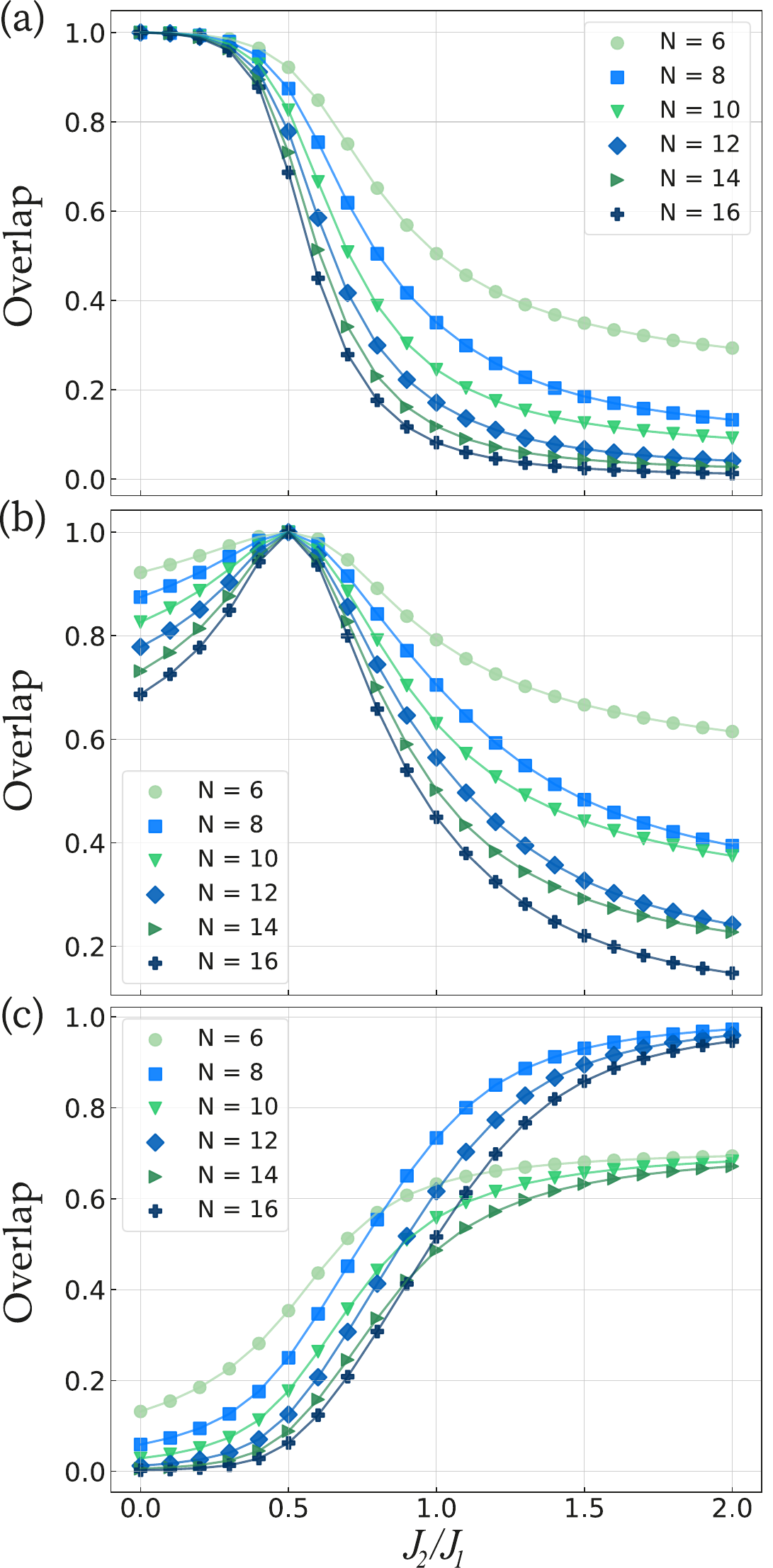}
  \caption{Overlap between the ground states of the $J_1 -J_2$ model calculated at different $J_2$ and three distinct solutions of the same model: (a) the eigenstate obtained at $J_1 = 1$ and $J_2 = 0$, (b) the eigenstate obtained at $J_1 = 1$ and $J_2 = 0.5$ and (c) the eigenstate obtained at $J_1 = 0$ and $J_2 = 1$. These calculations were done for the system with open boundary conditions.}
  \label{fig4}
\end{figure}

To explain the behaviour of the sign functions presented in Fig.\ref{fig3} we have estimated the overlap between exact diagonalization ground states, $\ket{\Psi^{\rm ED} (J_2)}$ for $J_{2} \in [0,2]$ and three distinct wave functions that can be positivized exactly. More specifically, these wave functions are ground states of the $J_1 - J_2$ Hamiltonian obtained at (i) $J_1 = 1$ and $J_2 = 0$, (ii) $J_1 = 1$ and $J_{2} =0.5$ (the Majumdar-Ghosh point) and (iii) $J_1 = 0$ and $J_{2} = 1$ that imitates the $J_{2} \rightarrow \infty$ regime. Figure \ref{fig4} (a) shows that the eigenstates from the range $J_2 \in [0,0.5]$ are characterized by a strong overlap ($> 0.6$) with $\ket{\Psi^{(\rm i)}}$ and $\ket{\Psi^{(\rm ii)}}$, which explains the existence of the sign function plateau at the maximal value in this parameter range. For the eigenstates from the range $J_2 \in (0.5,2]$ the overlap with the wave functions (i) and (ii) becomes smaller as $J_2$ increases, however, the contribution from $\ket{\Psi^{(\rm ii)}}$ to $\ket{\Psi^{\rm ED} (J_2)}$ at $J_2 >0.5$ decays more slowly than that from $\ket{\Psi^{(\rm i)}}$. At the same time, the overlap of exact diagonalization eigenstates with 
$\ket{\Psi^{(\rm iii)}}$ increases as $J_{2}$ approaches to 2 (Fig.\ref{fig4} (c)). Since both $\ket{\Psi^{(\rm ii)}}$ and $\ket{\Psi^{(\rm iii)}}$ are positivized exactly with the odd (even) protocols, the sign function features a minimum that is located at $J_2$ = 0.7 for the smallest systems considered here and shifts to $J_2 = 1$ as the system's size increases. 

The case when the system is characterized by odd $N_{\frac{1}{2}}$ requires a separate treatment, since the ground state of the $J_{1} - J_{2}$ model at $J_1 = 0$ and $J_2 =1$ is degenerate. For the considered systems of 6, 10, or 14 spins a fourfold degeneracy is observed. This means that instead of a single eigenstate there is a set of linearly independent wave functions, $\{\ket{\Psi^{(\rm iii)}_{n}} \}_{n = 1..4}$. To estimate the overlap in this case, we first find four real-valued linear combinations of $\{\ket{\Psi^{(\rm iii)}_{n}} \}_{n = 1..4}$ that are mutually orthogonal and each of which can be completely positivized with the odd protocol. Then at each $J_{2}$ we calculate the overlap between the positivized degenerate states and $\ket{\Psi^{\rm ED} (J_2) }$. The wave function from the degenerate set with the largest overlap is further used for analysis (Fig.\ref{fig4} (c)).

 
\begin{figure}[!t]
  \centering
  \includegraphics[width=0.93\linewidth]{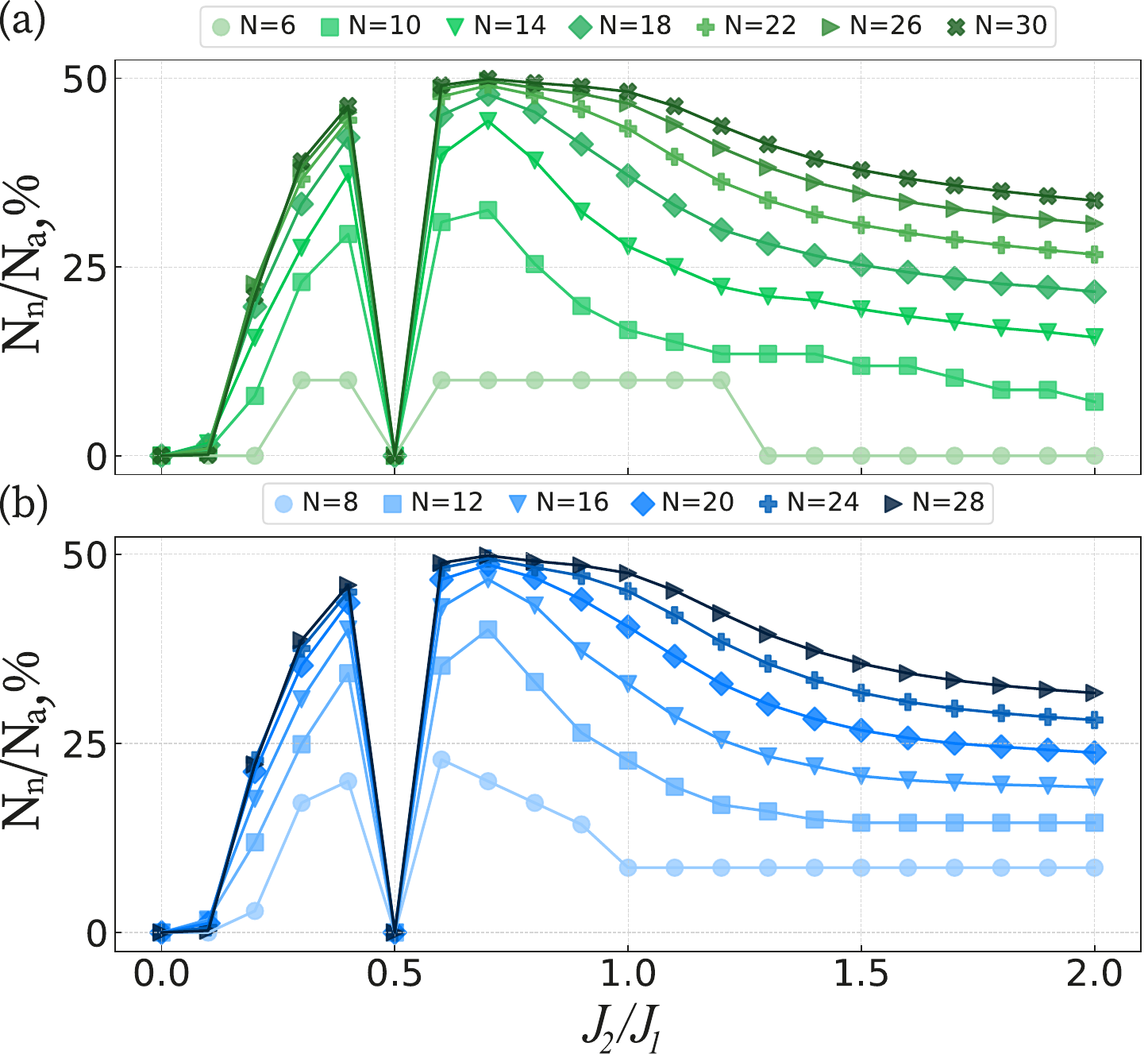}
  \caption{Dependence of the fraction of the basis states with negative amplitudes $N_n/N_a$ on $J_{2}$. (a) and (b) correspond to the cases of odd and even $N_{\frac{1}{2}}$, respectively.}
  \label{fig5}
\end{figure}

To complete our consideration of the quantum state sign structures we estimate the fraction of the basis functions with negative sign $N_n/N_a$ after performing quantum states positivization with one-qubit protocols. Here, $N_n$ is the number of negative amplitudes, $N_a$ is the total number of non-zero probability amplitudes. From Fig.\ref{fig5} one can see that the number of such basis states with negative amplitudes $N_n$ grows as the system's size increases for all the considered values of $J_2$ except $J_2 = 0$ and $J_2 = 0.5$ for which we get $\braket{\rm Sign}_{\psi} =1$. For the largest quantum systems we consider the fraction of negative basis states is close to 50\% below and above the critical $J_2 = 0.5$. This shows the main difference of the pozitivization results obtained with the Marshall-Peierls ($J_2 < 0.5$) and odd (even) ($J_2 > 0.5$) protocols.  In the former case the probabilities of the basis states with negative amplitudes are negligibly small in comparison to positive ones and practically independent of the system size. Considering $J_{2} > 0.5$ (the odd and even protocols), we observe a gradual enhancement of the negative contributions to the quantum states as the system size increases. 

The ultimate goal of employing the positivization procedure is to simplify the consideration of a target model by means of a unitary transformation, $U$ of the corresponding Hamiltonian. This leads to another spin Hamiltonian, $\hat {\widetilde H} = \hat U \hat H \hat U^{\dagger}$ whose ground states will be characterized by a smaller or zero fraction of the basis states with negative amplitudes. While the transformation corresponding to the Marshall-Peierls protocol can be found in the original paper, the odd and even schemes are constructed with the following unitaries:
\begin{equation}
 \hat U_{\rm odd/even} = \otimes_{j \in B} e^{i \pi \hat \sigma^z_{j} /2}, 
\end{equation}
where the sublattice B is selected in accordance with  Figs.\ref{fig1} (d) and \ref{fig2} (b).

\begin{figure}[!b]
  \centering
  \includegraphics[width=0.83\linewidth]{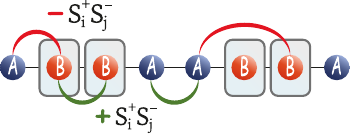}
  \caption{Schematic of sign structure of the Hamiltonian obtained with the even protocol. 
  Qubits are represented by blue and red circles, corresponding to the subsystem partitioning in Fig.\ref{fig1}(d) and rectangles indicate $R^z$ gate positions according to the rules shown in Fig.\ref{fig2}. The Hamiltonian terms that preserve spin-exchange operator signs are marked with green lines, the terms, for which the corresponding sign changes, are denoted with red lines.}.
  \label{fig6}
\end{figure}

The resulting Hamiltonian with positivized ground states is given by 
\begin{eqnarray}
\label{D OBC_n}
\hat {\widetilde H} =  &   J_1(\sum\limits_{i=1}^{N-1}\hat S_i^z \hat S_{i+1}^z + \frac{1}{2}\sum\limits_{i=1}^{\frac{N}{2}-1}\left(\hat S_{2i}^+ \hat S_{2i+1}^- + \right. 
 \left. \hat S_{2i}^- \hat S_{2i+1}^+\right) \nonumber \\ - & \frac{1}{2}\sum\limits_{i=1}^{N/2}\left(\hat S_{2i-1}^+ \hat S_{2i}^- + \hat S_{2i-1}^- \hat S_{2i}^+\right)) \nonumber  \\
 + & J_2\sum\limits_{\braket{\braket{i,j}}}\left(\hat S_i^z \hat S_j^z -\frac{1}{2}\left(\hat S_i^+ \hat S_j^- + \hat S_i^- \hat S_j^+\right)\right).
\end{eqnarray}
Rotation of the B sublattice spins around the z-axis by $\pi$ leaves the longitudinal interaction term $\hat{S}_i^z\hat{S}_j^z$ the same and flips the signs of the spin-exchange terms $\hat{S}^+_i \hat{S}^-_j$. For clarity in defining the Hamiltonian Eq.\ref{D OBC_n}, Fig. \ref{fig6} provides a graphical interpretation of sign assignments for corresponding terms. 

\subsection{Periodic boundary conditions}

The next step of our investigation is to describe the ground state properties of the one-dimensional $J_1-J_2$ model with periodic boundary conditions (PBC). As it was shown in Ref.\onlinecite{tonegawa} the PBC solutions obtained with exact diagonalization for small-size chains are characterized by two-fold degeneracy at the specific ratios between nearest and next-nearest neighbour exchange interactions. For instance, in the case of the 20-spin model there are degenerate ground states at $J_2=0.5, 0.521, 0.578, 0.703$ and 1.080. Such a degeneracy is due to the periodic boundaries, which imposes $N+1 \equiv 1$ and $N+2 \equiv 2$. Subsequent works \cite{Eggert, Sandvik} brought to conclusion that in the thermodynamics limit the degeneracy takes place for all $J_{2} > 0.24$, which calls one for the use of appropriate numerical methods \cite{Qskyrmion, Okatev}. In turn, according to Refs.\onlinecite{Becca, Torlai2018} the $J_1 -J_2$ model ground states can be distinguished with respect to the complexity of the sign structure. Depending on the particular value of next-nearest neighbour coupling, one can discriminate two different regions. The first one concerns the ground states obtained for $J_2 \in [0,0.5]$ for which the sign structure is trivial in a sense that it can be positivized with the Marshall-Peierls rule. Another parameter region $J_{2} > 0.5$ is characterized by non-trivial sign structure of the ground eigenfunctions that is almost insensitive to the unitary transformation with the MPR protocol, which represents the main interest for us in this study. 

\begin{figure}[!b]
  \centering
  \includegraphics[width=0.93\linewidth]{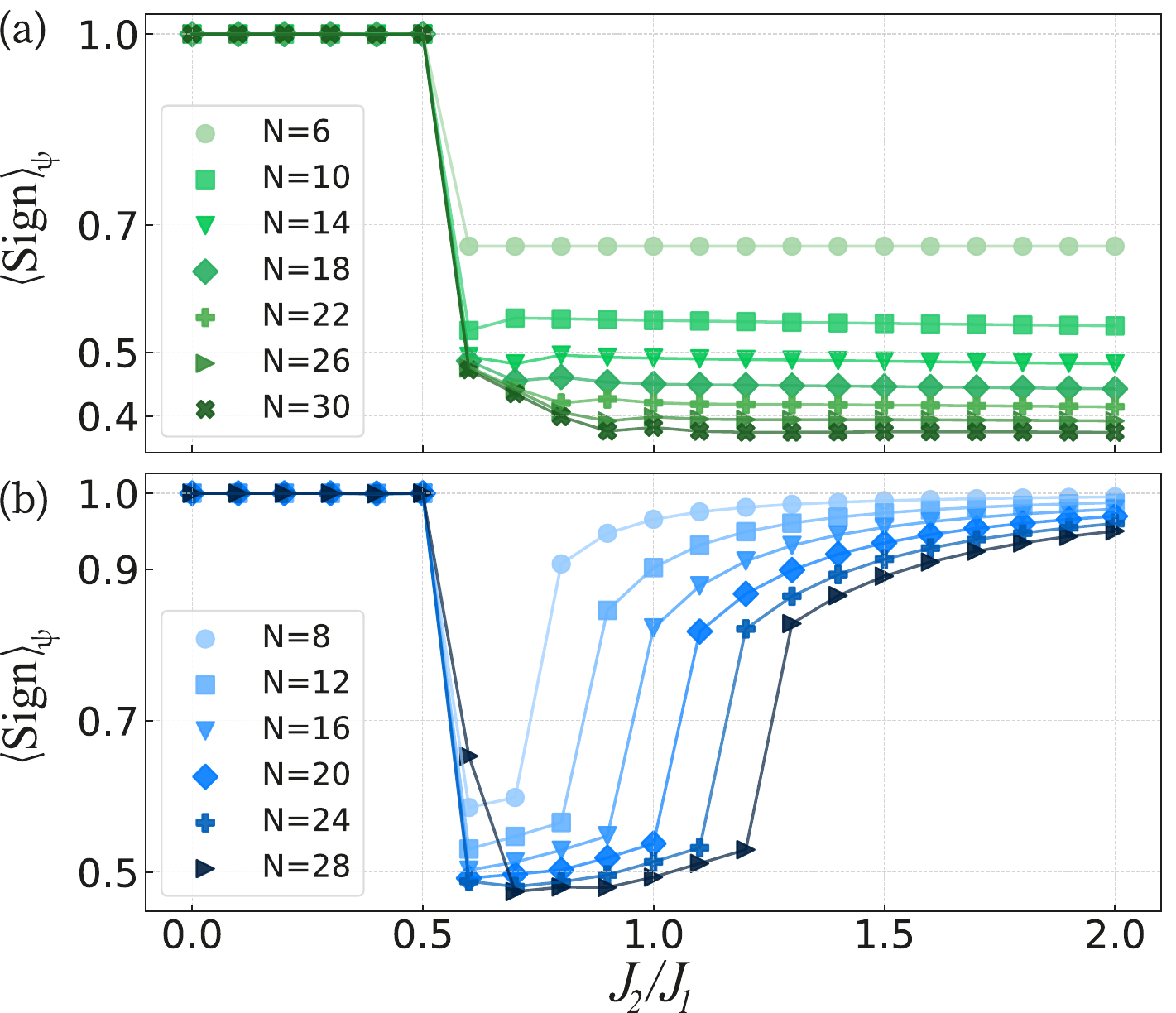}
  \caption{Estimates of the eigenfunction sign structure for systems with odd (a) and even (b) $N_{\frac{1}{2}}$. These results were obtained for the one-dimensional $J_1-J_2$ model with periodic boundary conditions.}
  \label{fig7}
\end{figure}

First, we reproduce the results of the previous works that study the sign structure of the system in question for the range of $J_2 \in [0,0.5]$. To be more concrete, Figs.\ref{fig7} (a) and (b) show that the sign function, Eq.\ref{metric} calculated at these values of $J_2$ is very close to 1 after applying the Marshall-Peierls rule. It is exactly equal to 1 for $J_2 = 0$ and $J_2 = 0.5$. In the latter case of the Majumdar-Ghosh point the system's ground state is two-fold degenerate for any chain length and the corresponding eigenfunctions can be defined as
\begin{eqnarray}
\ket{\Psi^{\rm 1}_{\rm MG}} = \ket{\Psi_{\rm S}}_{1,2}  \otimes \ket{\Psi_{\rm S}}_{3,4} \otimes ...\otimes \ket{\Psi_{\rm S}}_{N-1, N},
\end{eqnarray}
and 
\begin{eqnarray}
\ket{\Psi^{\rm 2}_{\rm MG}} = \ket{\Psi_{\rm S}}_{2,3}  \otimes \ket{\Psi_{\rm S}}_{4,5} \otimes ...\otimes \ket{\Psi_{\rm S}}_{N, 1}.
\end{eqnarray}
Similar to the OBC case, such a singlet-like structure of the ground state enables the complete positivization with the Marshall-Peierls transformation.

Considering the range $J_2 > 0.5$, we first examine behaviour of the sign function, Eq.\ref{metric} of the ground states transformed with the odd and even protocols found for positivization of the systems with open boundaries. In the PBC case the systems of odd and even $N_{\frac{1}{2}}$ are characterized by different behavior of $\braket{\rm Sign}_{\psi} (J_2)$. For odd $N_{\frac{1}{2}}$ the sign function reveals a constant profile whose value gradually decreases from 0.68 (N = 6) to 0.38 (N = 30), which can be classified as a partial positivization with respect to the initial eigenfunctions before the unitary transformation. In the case of the systems with even $N_{\frac{1}{2}}$, the $\braket{\rm Sign}_{\psi}$ function features a pronounced minimum and approaches to 1 at $J_{2}/J_{1} \rightarrow 2$. 


\begin{figure}[t!]
\begin{minipage}[h]{1\linewidth}
\center{\includegraphics[width=1\linewidth]{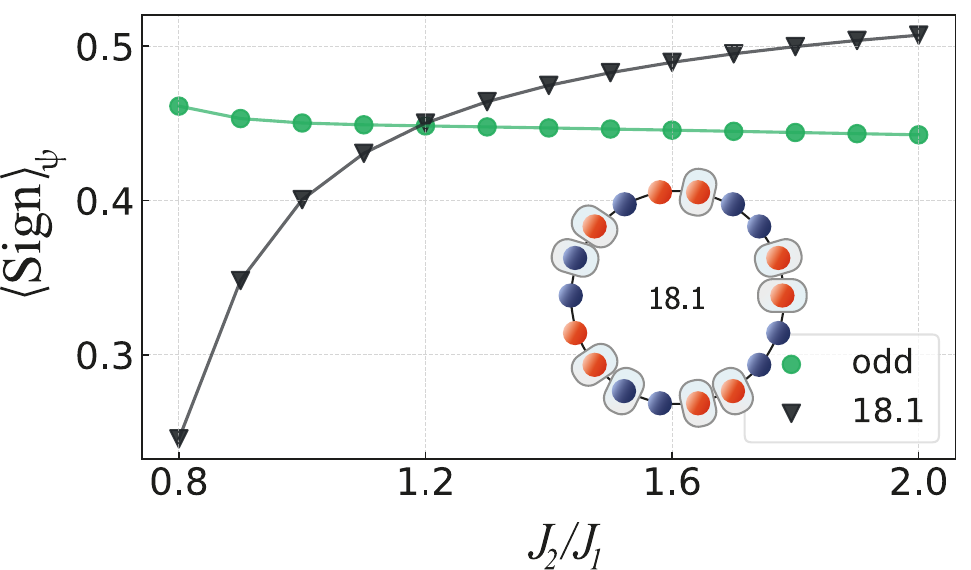}}
    \caption{Comparison of sign function values obtained with odd (green) and 18.1 (black) protocols applied to an 18-spin chain with periodic boundary conditions. Inset gives a schematic of the 18.1 scheme. Shaded ovals denote qubits whose states are modified with $R^z(\pi)$ gates.}
    \label{fig8}
    \end{minipage}
\end{figure}

We have also found that in the case of odd $N_{\frac{1}{2}}$  one can employ different protocols that slightly improve the results obtained with the odd scheme. They can be revealed with brute-force search described in the methodological part. An example of the thus found unitary transformations is visualized in Fig.\ref{fig8} for an 18-spin chain with PBC, shown in black.
One can see that the 18.1 positivization protocol outperforms the odd one for $J_2 > 1.2$
In addition, we have concluded that the Marshall-Peierls protocol is more effective for the PBC chains of 10, 12, 14, 16, 18, 24, 26, 30 spins at $J_2=0.6$ and for 14-spin system at $J_2=0.7$. 
   
Thus, in our study we provide a comprehensive approach to positivization problem of the one-dimensional $J_1 - J_2$ model ground states with different $R^z$ gate protocols in the wide range of parameters. This contrasts to previous works where the authors either employ the Marshall-Peierls rule for different $J_2$ values at the fixed bipartition (Fig.\ref{fig1} (c)) or optimized the sign structure for the particular $J_2/J_1$ ratio as it was done in Ref.\onlinecite{Torlai_positivization}. The protocol found in the latter work for $J_{2}/J_{1} = 2$ assumes rotations on spins from both sublattices with angles separated by $\pi$ (Fig.\ref{fig2} (c)). For the same setting we propose the even protocol that is characterized by half the number of $R^z$ gates than those in Ref.\onlinecite{Torlai_positivization}. 

In the next section we discuss protocol based on two-qubit gates that enables a further pozitivization of the $J_1-J_2$ ground states. However, this complicates the consideration and requires one to the need to consider more complex Hamiltonians with multi-spin interactions. 

\subsection{Pozitivization with two-qubit CZ gates}

\begin{figure}[ht]
\centering\includegraphics[width=0.15\textwidth]{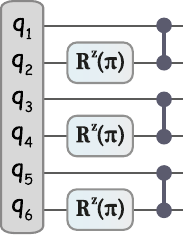}
    \caption{6-qubit example of the MPR+CZ positivization protocol combining layers of one-qubit $R_z$ gates and two-qubit CZ gates.}
    \label{fig9}
\end{figure}

Since the considered protocols to positivize sign structure with single-qubit gates demonstrate low efficiency for chain of odd $N_{\frac{1}{2}}$ with periodic boundary conditions in the region $J_2>0.5$ (Fig. \ref{fig7}), it is instructive to analyze effect of applying the two-qubit CZ gates on the sign function of the ground state solutions. For that we performed the brute-force search for particular combinations of the CZ gates that decrease the contribution of the basis states with negative amplitudes. Joint schemes combining the CZ-based protocols with Marshall-Peierls and odd (even) ones have been also examined. It was found that the best positivization results for $N$-qubit system are achieved when combining the transformation based on Marshall-Peierls rule with a set of $N/2$ CZ gates acting on qubit pairs without overlap as exemplified in Fig.\ref{fig9}.    

\begin{figure}[b]
     \centering
    \includegraphics[width=0.5\textwidth]{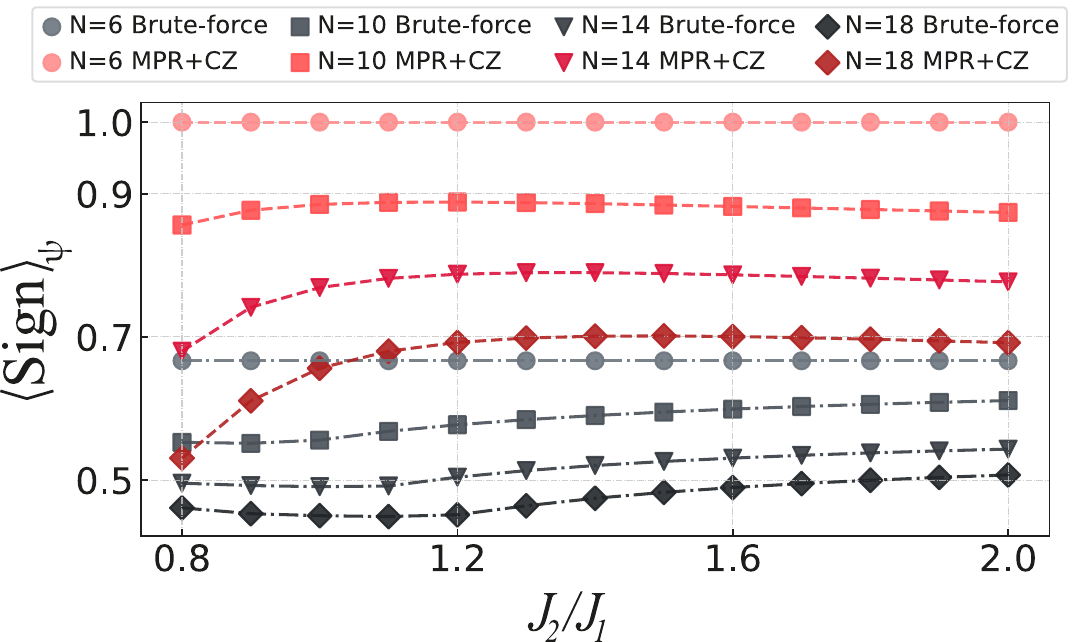}
    \caption{Comparison of positivization results for systems with odd $N_{\frac{1}{2}}$ by using MPR+CZ scheme (shades of red) and single-qubit protocols (shades of gray) found with brute-force procedure.}
    \label{fig10}
\end{figure}

As follows from Fig. \ref{fig10}, in the case of the 6-spin system the combined protocol produces the quantum states for which the sign function is exactly equal to 1 in the range $J_2 \in [0.8, 2]$. It also considerably improves positivization results with respect to those obtained with single-qubit odd protocols for $N = 10, 14$ and 18. For instance, at $J_2 = 1$ the sign function changes from 0.45 (odd protocol) to 0.66 (MPR+CZ) for 18-qubit system.

\begin{figure}[ht]
     \centering
\includegraphics[width=0.42\textwidth]{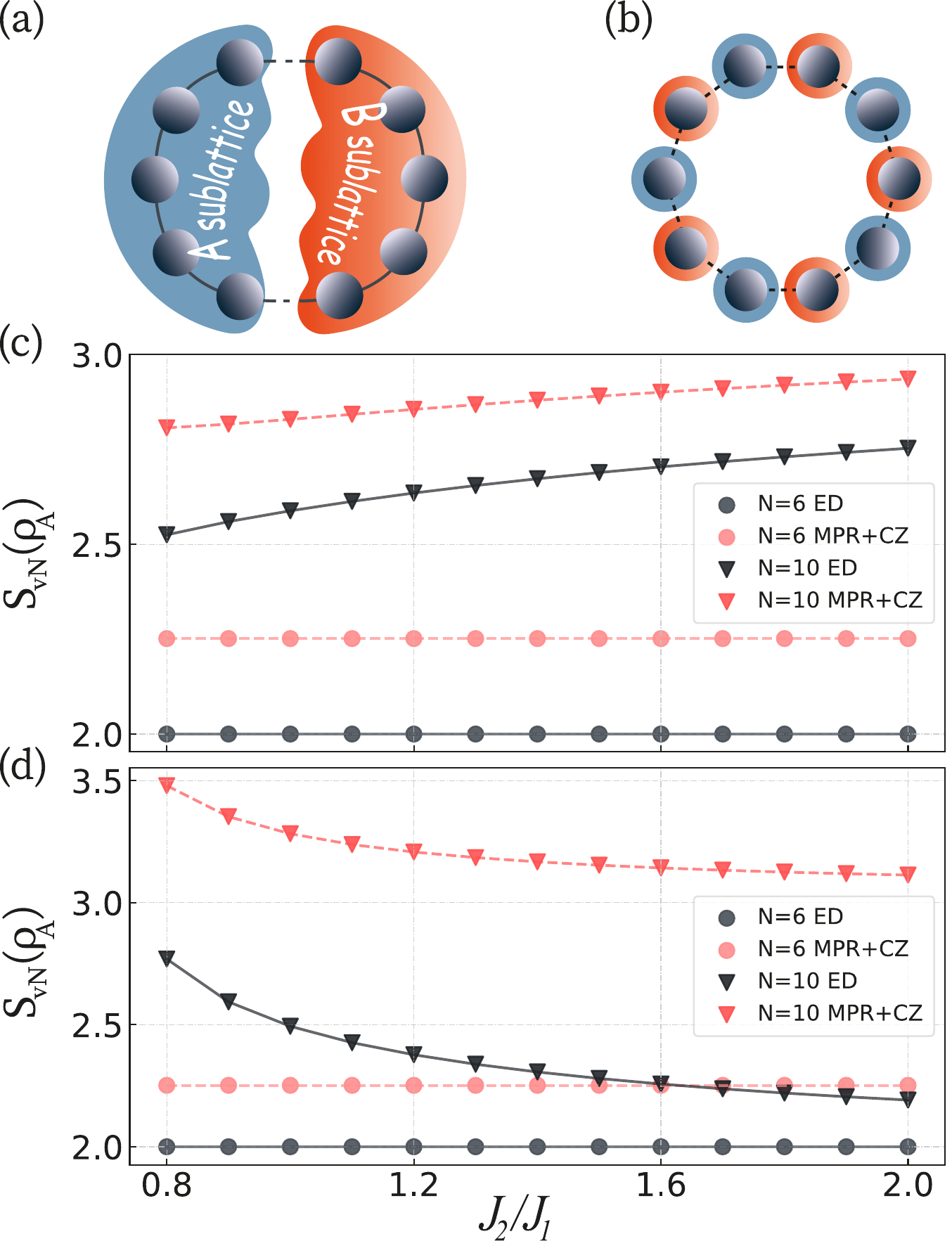}
    \caption{Calculation of the entanglement entropy. (a) and (b) panels visualize the bipartitions used to define the reduced density matrix. (c) and (d) panels demonstrate the entropies calculated with the bipartitions presented in (a) and (b), respectively. The entanglement entropy of the states obtained from exact diagonalization are shown in black and the entropy of the states modified with the MPR+CZ protocol \--- pink colour.}
    \label{fig11}
\end{figure}

While the single-qubit transformations (Marshall-Peierls, odd or others) don't affect the quantum correlations in the systems in question, it is not the case for the MPR+CZ protocol. To probe the corresponding changes in quantum correlations when imposing different sign structures we have calculated the von Neumann entanglement entropy, $S_{\rm vN} (\rho_{\rm A}) = - {\rm Tr} \rho_{\rm A} \log_{2} \rho_{\rm A}$ for the reduced density matrix $\rho_{\rm A} = {\rm Tr}_{\rm B} \rho_{\rm AB}$. For that bipartitions of two types presented in Figs.\ref{fig11} (a) and (b) were used. The examples of 6- and 10-spin systems indicate a considerable amplification of the entanglement entropy after applying the joint MPR+CZ protocol.  

While the implementation of two-qubit CZ gates considerably improves the positivization results, using such a procedure for finding ground state of the $J_1 - J_2$ model, Eq.\ref{Hamiltonian} is complicated by the fact that one needs to solve a multi-spin Hamiltonian. Indeed, taking the unitary transformation $\hat U=\prod\limits_i^{N/2}[{\rm I}_{2i-1}\otimes R_{2i}^z(\pi)]{\rm CZ}_{2i-1,2i}$ corresponding to the MPR+CZ protocol (Fig.\ref{fig9}), one can represent the CZ gate between $i$th and $j$th qubits in the form \cite{Dicke_sign}
\begin{equation}\label{CZ_equation}
CZ_{i,j} = \frac{I_i\otimes I_j + \sigma_i^z\otimes I_j + I_j
\otimes\sigma_j^z-\sigma_i^z\otimes\sigma_j^z}{2}.
\end{equation}
This gives the following Hamiltonian 
\begin{equation}\label{CZ_Hamiltonian}
    \begin{array}{ll}
    \hat H' = \hat U \hat H \hat U^{\dagger} = & \\
    J_1\left( \sum\limits_{\braket{ij}} \hat S_{i}^z \hat S_{j}^z - \sum\limits_{i}^{N/2} \sum\limits_{\mu} (\hat S_{2i-1}^{\mu} \hat S_{2i}^{\mu} +\frac{1}{4} \hat S_{2i- 1}^z \hat S_{2i}^{\mu} \hat S_{2i+1}^{\mu} \hat S_{2i+2}^z)\right) &\\
    +  J_2\left(\sum\limits_{\braket{\braket{ij}}} \hat S_i^z \hat S_{j}^z + \frac{1}{4} \sum\limits_i^{N/2} \sum\limits_{\mu} (\hat S_{2i-1}^{\mu} \hat S_{2i}^z \hat S_{2i+1}^{\mu} \hat S_{2i+2}^z \right. &\\
    + \left. \hat S_{2i-1}^z \hat S_{2i}^{\mu} \hat S_{2i+1}^z \hat S_{2i+2}^{\mu})\vphantom{\sum\limits_i^N}\right),&\\
    \end{array}
\end{equation}
where $\mu=\{x,y\}, N+1\equiv1, N+2\equiv2$. The obtained model features four-spin interactions and further positivization steps that can be done by using multi-qubit controlled-Z gates will lead to even more complicated models. Thus, a choice must be made between the complexity of the Hamiltonian components and the simplicity of the eigenstate sign structure.

\section{Conclusion and outlook}
To sum up, we have performed a systematic study of the sign structure of the ground states of the one-dimensional $J_1 -J_2$ model. Our contribution is the following. In addition to the famous Marshall-Peierls sign rule that allows positivization of the ground states in the range $J_{2} / J_{1} \in [0, 0.5]$,  unitary transformations producing the best positivization results for $J_{2} / J_{1} \in (0.5,2]$ have been found. The sensitivity of the results obtained with a particular positivization scheme to the parity of $N/2$ and boundary conditions has been demonstrated. For the parameter ranges characterized by a low-level positivization we have additionally considered distinct schemes that combine unitary transformations based on the Marshall-Peierls sign rule and two-qubit CZ gates.  

The protocols we found can be primarily used to improve the  solutions of the one-dimensional $J_1 - J_2$ model obtained within neural quantum state approach. In particular, the results reported in Refs.\onlinecite{Shamim, Becca} evidence that encoding MPR structure into the Hamiltonian used for variational learning or attaching the Marshall-Peierls sign rule to the wave-function amplitudes of a RBM representation provide a clear computational advantage, since they improve the estimates of the ground state energies and make the optimization of the variational state easier. Interestingly, a positive effect of accounting the sign structure has been revealed even for the cases with $J_{2} >0.5$ where the MPR doesn't work. Thus, our findings suggest further improvement of these neural network approaches by using the odd and even protocols to positivize the ground states.

\section{Acknowledgments}
This work was supported by the Ministry of Science and Higher Education of the Russian Federation (theme FEUZ-2026-0010).

\end{document}